\documentstyle[eqsecnum,aps]{revtex}
\begin{document}
%
%
%
%
%
%
\title{
Unified Description of Freeze-Out Parameters in Relativistic Heavy Ion
Collisions.}
\author{J. Cleymans$^1$ and   K.~Redlich$^{2,3}$}
\address{
$^1$Department of Physics, University of Cape Town, Rondebosch 7700,\
South Africa  \\
$^2$Gesellschaft f\"ur Schwerionenforschung (GSI),
D-64220 Darmstadt, Germany\\
$^3$Department of Theoretical Physics, University of Wroclaw, Wroclaw,
Poland\\}
\maketitle

\begin{abstract}
It is shown that the chemical freeze-out parameters obtained at  CERN/SPS,
BNL/AGS  and  GSI/SIS energies all correspond to a unique value
of  1  GeV  per  hadron  in the local rest frame of the system,
independent  of  the  beam  energy  and  of the target and beam
particles.
\end{abstract}
We  have  found  that  a  unified  description  of the hadronic
abundances  produced  in  heavy ion collisions at the CERN/SPS,
the  BNL/AGS  and  the  GSI/SIS  accelerators is possible. This
description   covers   a   range  in  beam  energies  from  200
A$\cdot$GeV  to  below  1  A$\cdot$GeV.  As  it  turns  out the
same description can also be applied to the hadronic abundances
in  LEP  and  in  $p-p$ and $\bar{p}-p$ collisions
with slightly  different treatment of strangeness sector which
accounts for strangeness undersaturation.
 The
result
can be summarized in a surprisingly simple way: the
hadronic  composition  of  the  final state is determined by an
energy  per  hadron being approximately 1 GeV per hadron in the
rest  frame  of  the  produced  system.  This  generalizes  an
observation made a long time ago by Hagedorn \cite{hagedorn}
 for  $p-p$
collisions, namely, as one increases the
beam energy, the available energy is used to produce more
particles, but not to increase the temperature of the system.  
This  led  Hagedorn  to  the  idea  of  a  limiting
temperature.  For  heavy  ion collisions one has to take into
account  not only the temperature but also 
the finite baryon density
of  the  system,  which is described  by  the  baryon 
chemical potential
$\mu_B$.
This, as it was first indicated by P. Braun-Munzinger and J. Stachel 
\cite{stachel},
 leads to a freeze-out curve in the $T,\mu_B$ plane. In
Fig. 1  the  values  of the freeze-out parameters are shown, as
obtained   by  various  groups  (a  recent summary  can  be  found  in
 \cite{sollfrank}). 
The solid line corresponds to 1 GeV
per hadron, the dashed line corresponds to 0.94 GeV per hadron.
This  energy  corresponds  to  the  chemical  freeze-out stage,
namely,  before the hadrons  decay into 
the stable hadrons.

It is well known that 
various
effects (e.g.  flow) severely distort 
the momentum
spectra of the particles produced in heavy ion collisions. 
It has been repeatedly pointed out however that many
of these effects  cancel out  for 
ratios of fully integrated 
particle multiplicities (see e.g. \cite{jaipur}). 
The analysis of particle ratios 
is  therefore the best method to obtain reliable information on
the chemical freeze-out parameters of the hadronic final state.
Such  an  analysis,  relying  as  much  as  possible  on  fully
integrated   particle   multiplicities   was   carried   out
 for BNL/AGS and for CERN/SPS data. 
 In  Fig.   1  the  SPS  points  are indicated by open squares
 \cite{PBM,bgs}  while  the  AGS  points  are indicated by open
 circles \cite{PBM,cets,bgs}.

Data  using $Ni$ and $Au$ beams at energies between 0.8 and 1.9
A$\cdot$GeV  have  become  available  recently from the GSI/SIS
accelerator.  These  data  have attracted considerable interest
due  to  the  surprisingly  large  number of $K^-$ mesons being
produced   below  threshold.  A  very  detailed  and  extensive
discussion of these results in the framework of thermal models 
has been presented in \cite{ceks,averbeck,oeschler}. 
The results for the
freeze-out  parameters  
for  $Ni-Ni$  at  1.9  A$\cdot$GeV is shown as an open triangle
in Fig.  1.
The
points  with  the  lowest  temperature  correspond  to  $Au-Au$
collisions at 0.8 and 1.0 A$\cdot$GeV and $Ni-Ni$ collisions at
1.0 and 1.8 A$\cdot$GeV. 
and are also shown as open triangles.

A similar   analysis   has  been  performed  in
\cite{becattini} for $e^+e^-$
annihilation into hadrons at LEP.
Since  no  baryons  are  involved here this corresponds to zero
baryon chemical potential, $\mu_B=0$. An impressive fit has
been  obtained  here  since  no less than 29 different hadronic
abundances  can  be reproduced. It is our view that such a good
agreement  cannot  simply  be  a coincidence. This analysis was
subsequently   extended \cite{becattini} to  $p-p$  and
$\bar{p}-p$  reactions  at  CERN. 
 In  this  case, 
one reproduces the Hagedorn temperature obtained many
years ago.

In the underlying hadronic gas model all these points 
can be described by a single
curve corresponding to a fixed energy per particle, $\epsilon/n$, 
which has
approximately the value of 1 GeV per particle in the hadronic gas. 
{\it This value characterizes all the final states produced by 
beams having
1  A$\cdot$GeV  all  the  way  up  to  200   A$\cdot$GeV.}
Thus, the only modification one needs to make to the concept
of Hagedorn's limiting temperature it that there exist
a "limiting" - freeze-out energy per particle of 1 GeV at which
hadrons are formed in a collision. 

This  observation  leads to a considerable
unification  in  the  description  of  the
hadronic final states produced in high 
energy collisions.
 
\acknowledgments{We acknowledge 
stimulating  discussions with P. Braun-Munzinger, H. Satz, H.
Specht and J. Stachel.
The help of Helmut Oeschler with
the  GSI/SIS  data  was  essential  in  compiling  Fig.  1. We
acknowledge  also  the  generous hospitality of the theoretical
physics division of the GSI.}


%
%
%
%
%
\begin{figure}
\caption{
Freeze-out   values   obtained   from  hadronic  abundances  at
CERN/SPS,  BNL/AGS  and  GSI/SIS. Also indicated are the points
obtained  from observed hadronic abundances at LEP and in $p-p$
collisions  at  CERN.  The  smooth curves correspond to a fixed
energy per hadron in the hadronic gas model. 
}
\end{figure}
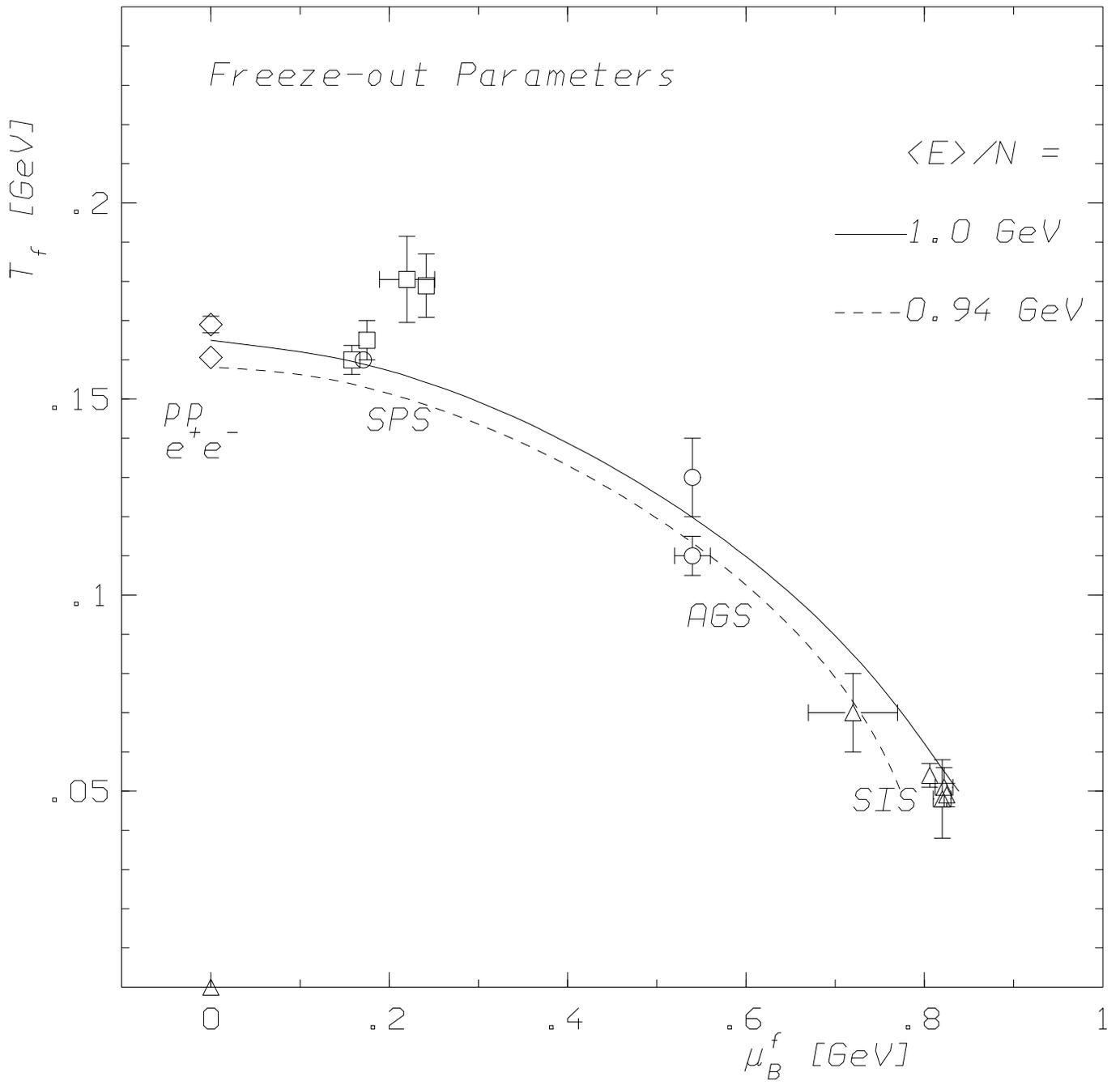
\end{document}